\documentclass[preprint,prd,amsmath,amssymb,nofootinbib,tightenlines
,floatfix,letterpaper]{revtex4-1}
\usepackage[dvips]{graphics}
\usepackage{epsfig}
\usepackage{latexsym}
\usepackage{slashed}
\newcommand{\be}{\begin{equation}}
\newcommand{\ee}{\end{equation}}
\newcommand{\bea}{\begin{eqnarray}}
\newcommand{\eea}{\end{eqnarray}}

\begin{document}
\preprint{ NUHEP-TH/13-03}
\preprint{FERMILAB-PUB-13-212-T}

\title{$H\rightarrow \gamma\gamma$ as a Triangle Anomaly:  Possible Implications for the Hierarchy Problem}

\author{Andr\'{e} de Gouv\^{e}a\footnote{Electronic address:  degouvea@northwestern.edu}
}
\affiliation{Department of Physics and Astronomy, Northwestern University, Evanston, IL 60208 USA}

\author{Jennifer Kile\footnote{
	Electronic address: jenkile@northwestern.edu}
}
\affiliation{Department of Physics and Astronomy, Northwestern University, Evanston, IL 60208 USA}

\author{Roberto Vega-Morales\footnote{
    Electronic address:  robertovegamorales2010@u.northwestern.edu}
}
\affiliation{Department of Physics and Astronomy, Northwestern University, Evanston, IL 60208 USA\\and\\Fermi National Accelerator Laboratory (FNAL), Batavia, IL 60510-0500 USA}

\begin{abstract}
The Standard Model calculation of $H\rightarrow\gamma\gamma$ has the curious feature of being finite but regulator-dependent.  While dimensional regularization yields a result which respects the electromagnetic Ward identities, additional terms which violate gauge invariance arise if the calculation is done setting $d=4$.  This discrepancy between the $d=4-\epsilon$ and $d=4$ results is recognized as a true ambiguity which must be resolved using physics input; as dimensional regularization respects gauge invariance, the $d=4-\epsilon$ calculation is accepted as the correct SM result.  However, here we point out another possibility; working in analogy with the gauge chiral anomaly, we note that it is possible that the individual diagrams do violate the electromagnetic Ward identities, but that the gauge-invariance-violating terms cancel when all contributions to $H\rightarrow\gamma\gamma$, both from the SM and from new physics, are included.  We thus examine the consequences of the hypothesis that the $d=4$ calculation is valid, but that such a cancellation occurs.  We work in general renormalizable gauge, thus avoiding issues with momentum routing ambiguities.  We point out that the gauge-invariance-violating terms in $d=4$ arise not just for the diagram containing a SM $W^{\pm}$ boson, but also for general fermion and scalar loops, and relate these terms to a lack of shift invariance in Higgs tadpole diagrams.  We then derive the analogue of ``anomaly cancellation conditions'', and find consequences for solutions to the hierarchy problem.  In particular, we find that supersymmetry obeys these conditions, even if it is softly broken at an arbitrarily high scale.
\end{abstract}

\maketitle

\section{Introduction}
\label{sec:intro}
The process $H\rightarrow\gamma\gamma$ has great potential to reveal the effects of new physics.  This decay is absent at tree level but arises at one loop; in the Standard Model (SM) its main contribution comes from a $W$-boson loop, with effects from the top quark also playing a significant role.  However, notably, $H\rightarrow\gamma\gamma$ is also sensitive to heavy charged particles arising in physics beyond the SM.  For this reason, a precise measurement of the $H\rightarrow\gamma\gamma$ branching fraction is highly anticipated.  Currently, the branching fraction of the Higgslike particle discovered at LHC to $\gamma\gamma$ is measured to be $0.78\pm 0.27$ (CMS) \cite{CMS:2013}  and $1.65\pm0.24\mbox{(stat)}^{+0.25}_{-0.18}\mbox{(syst)}$ (ATLAS) \cite{ATLAS:2013oma} times the SM prediction. 

The SM prediction for $H\rightarrow\gamma\gamma$ at one-loop order has been known for some time \cite{Ellis:1975ap,Shifman:1979eb,Rizzo:1979mf,Ioffe:1976sd,Resnick:1973vg}.  As noted long ago \cite{fukuda1:1949,fukuda2:1949,fukuda3:1949,Steinberger:1949wx,Gerstein:1969cx,Schwinger:1951nm}, the theoretical prediction of $H\rightarrow\gamma\gamma$ contains some peculiarities stemming from the regulator-dependence of momentum integrals which arise in intermediate steps of the calculation.  Although the final result for $H\rightarrow\gamma\gamma$ is finite, this comes about after a cancellation of logarithmically divergent momentum integrals\footnote{In unitary gauge, greater-than-logarithmically-divergent integrals arise; as discussed below, we will not use unitary gauge in this work.}; depending on the regulator used, there may or may not be leftover finite terms which affect the electromagnetic gauge invariance of the calculation.  This is an example of behavior pointed out in \cite{Jackiw:1999qq} which showed that certain finite integrals are regulator-dependent and argued that this regulator dependence constitutes a true ambiguity in the calculation which must be resolved using physics input.  In the SM, the use of dimensional regularization using dimension $d=4-\epsilon$ yields a result that obeys the electromagnetic Ward identities, while calculations using $d=4$ yield spurious terms which violate electromagnetic gauge invariance.  Thus, dimensional regularization is commonly used for this calculation; see, for example \cite{Ellis:1975ap,Rizzo:1979mf}. 

In this paper, we consider the possibility that the regulator dependence in $H\rightarrow\gamma\gamma$ is more than a mathematical curiosity.  Our motivations for this are as follows.  First, as  $H\rightarrow\gamma\gamma$ is finite and contains at most logarithmically-divergent terms, one might find it surprising that we lose gauge invariance simply in passing from $d=4-\epsilon$ to $d=4$ dimensions.  For this reason, we wish to explore the consequences of using 4-dimensional Lorentz invariance instead of gauge invariance as the physics input to determine the value of the ambiguous integrals in $H\rightarrow\gamma\gamma$.  In doing so, we will not abandon gauge invariance, but will instead require that the gauge-invariance-violating terms cancel when all contributions, from the SM and from new physics, to $H\rightarrow\gamma\gamma$ are included, analogous to the cancellation of symmetry-violating terms in the SM chiral anomaly.  We will discuss the plausibility of such a cancellation below.

Second, we wish to point out the simple but nonobvious fact that it is {\em possible} that the $d=4$ calculation of $H\rightarrow\gamma\gamma$ is valid.  In the scenario we outline here, the gauge-invariance-violating terms in the $d=4$ calculation are residual artifacts of not knowing the full theory contributions to  $H\rightarrow\gamma\gamma$.  If such a scenario is realized in nature, calculations done in the full theory would give the same result, whether calculated in $d=4$ or $d=4-\epsilon$ dimensions.  However, when the full theory is unknown, gauge-invariant regulators, by removing gauge-invariance-violating terms, may discard clues about new physics.  Thus, we wish to consider the possibility that these terms contain information about physics beyond the SM.  We will try to make this second point more precise below, after we review a few technical details of our analysis.  

We note that there have been many recent papers which have noted the regulator-dependence of the $H\rightarrow\gamma\gamma$ calculation.  The authors of \cite{Gastmans:2011wh,Gastmans:2011ks} used unitary gauge and $d=4$, concluding that the previous results on the gauge boson loop contribution to $H\rightarrow\gamma\gamma$ \cite{Ellis:1975ap,Shifman:1979eb,Rizzo:1979mf,Ioffe:1976sd,Resnick:1973vg} were in error.  (For responses to their work, see \cite{Shifman:2011ri,Huang:2011yf,Marciano:2011gm,Jegerlehner:2011jm,Shao:2011wx,Campbell:2011iw,Bursa:2011aa,Piccinini:2011az,Dedes:2012hf,Donati:2013iya,Cherchiglia:2012zp}.)  In this work, we use renormalizable gauge, with general $\xi$, but do not consider the case of unitary gauge, $\xi\rightarrow\infty$, which is plagued by greater-than-logarithmic ultraviolet divergences.  By choosing not to use unitary gauge, we will encounter only logarithmically-divergent and finite integrals in our calculation of $H\rightarrow\gamma\gamma$, and, as these integrals are invariant under a shift of the loop momentum, we do not need to worry about the details of relative momentum routing between diagrams.  Additionally, our regulator-dependent terms will be far simpler than those in \cite{Gastmans:2011wh,Gastmans:2011ks}; they will be limited to the finite constant term that has been Dyson-subtracted \cite{Dyson:1949ha} in some previous works \cite{Shao:2011wx,Piccinini:2011az}. (Also, see \cite{Huang:2011yf} for comments questioning the use of the Dyson subtraction; we also note that it is argued in \cite{Gerstein:1969cx} that, in this case, such a subtraction and using a Pauli-Villars regulator are equivalent.)  We note that there have been several recent works introducing new regulators or techniques to obtain gauge-invariant results for $H\rightarrow\gamma\gamma$ in $d=4$ \cite{Shao:2011wx,Donati:2013iya,Cherchiglia:2012zp,Huang:2011yf}; for a similar work using a momentum cutoff while preserving shift invariance, see \cite{Cynolter:2010ei}.  We do not adopt such a strategy of developing new gauge-invariant $d=4$ regulators here, but instead require that gauge invariance be recovered via a cancellation of the offending terms.

Although a cancellation between SM and new physics contributions which conspires to preserve gauge invariance in the $d=4$ calculation of $H\rightarrow\gamma\gamma$ may seem unlikely, we argue here that such a cancellation is not implausible.  First, we point out that the gauge-invariance-violating terms which arise in $d=4$ occur not just for the dominant $W^{\pm}$ loop, but also for general fermion and scalar loops.  Additionally, these terms are always of the same form; the contributions to the matrix element ${\cal M}_{\mu\nu}$ are all $\sim g_{\mu\nu}$.  Lastly, as we show in Secs. \ref{sec:fands} and \ref{sec:wlp}, the problematic terms in $d=4$ are related to shifts of quadratically-divergent Higgs tadpole diagrams, which are manifestations of the SM hierarchy problem.  (Here, when we refer to the hierarchy problem, we refer to the sensitivity of the Higgs mass-squared to quadratically-divergent radiative corrections.)  Thus, we have other theoretical reasons to suspect that such a cancellation may occur in models which address the problem of naturalness in the Higgs sector.  In Sec. \ref{sec:hier}, we will make the relation between the regulator dependence in $H\rightarrow\gamma\gamma$ and the hierarchy problem more precise; we find, for example, that the gauge-invariance-violating terms in the $d=4$ calculation do, in fact, cancel in supersymmetry, even if it is softly broken.

We briefly mention some additional motivation for suggesting a possible analogy between triangle anomalies and the $d=4$ calculation of $H\rightarrow\gamma\gamma$.  First is the observation that, in the case of the SM triangle anomalies, the quantum numbers of the SM fermions conspire to enforce the Ward identites of the SM gauge group.  This indicates that it is, in principle, possible that one or more symmetries may be preserved by the adjustment of the particle content of whatever constitutes the full theory of nature.  Furthermore, the two cases share a few additional similarities.  In both the triangle anomaly and the $d=4$ calculation of $H\rightarrow\gamma\gamma$, application of the Ward identity to one of the external bosons results in expressions containing linearly-divergent integrals; in both cases, the behavior of these integrals under loop momentum shifts plays a significant role in the loss of gauge invariance in the amplitude.\footnote{It has been pointed out to us by W. Marciano that there may be a relation between $H\rightarrow\gamma\gamma$ and the trace anomaly.  As the trace anomaly calculation requires handling quadratically-divergent vacuum polarization diagrams, we do not attempt to consider this possibility in $d=4$.}  One significant difference, however, between the SM chiral anomaly and the $d=4$ calculation of $H\rightarrow\gamma\gamma$ is that, unlike the case of the chiral anomaly, regulators which preserve gauge invariance in $H\rightarrow\gamma\gamma$ do exist.  Thus, unless we prioritize 4-dimensional Lorentz invariance over gauge invariance when choosing a regulator, no principle tells us that gauge invariance in $H\rightarrow\gamma\gamma$ must be enforced by a cancellation between diagrams.  However, if nature does enforce gauge invariance via such a cancellation, this places constraints on the particle content of physics beyond the SM.  Thus, the question of whether gauge invariance is enforced in $H\rightarrow\gamma\gamma$ via regularization of individual diagrams or a cancellation between diagrams can, in principle, be answered empirically.  We will discuss this further when we consider the relevance to the hierarchy problem in Sec. \ref{sec:hier}.    

While we will make the connection between the regularization ambiguities in $H\rightarrow\gamma\gamma$ and the hierarchy problem more precise in Sec. \ref{sec:hier}, we briefly sketch the argument here.  We first point out that in discussing the hierarchy problem, we are only referring to the sensitivity of the Higgs self-energy to quadratically-divergent radiative corrections; our results are not relevant for finite contributions arising from high-scale physics.  If we apply the Ward identity to both final-state photons in $H\rightarrow\gamma\gamma$, we obtain expressions which are closely related to momentum shifts of the loops in $H\rightarrow\gamma\gamma$ with both photons removed; these resulting tadpole diagrams are a subset of the diagrams which give quadratically-divergent contributions to the Higgs vacuum expectation value.   The difference in the momentum integral in going from $d=4-\epsilon$ to $d=4$ then arises due to differences in shift-invariance between linearly divergent and less-than-linearly divergent integrals.  Thus, when we derive the analogue of ``anomaly cancellation conditions'' for $H\rightarrow\gamma\gamma$, we are deriving the conditions under which these shift-invariance-violating terms cancel; we are taking a feature of dimensional regularization--invariance under momentum shifts--and, instead of enforcing it with a regulator, building it directly into those amplitudes relevant when the Ward identities are applied to $H\rightarrow\gamma\gamma$.

The remainder of this paper is organized as follows.  In Sec. \ref{sec:fands}, we consider the contributions to $H\rightarrow \gamma\gamma$ mediated by fermion and scalar loops; we discuss the expressions obtained by applying the Ward identity to these processes and the differences that arise depending on whether these expressions are evaluated in $d=4$ or $d=4-\epsilon$ dimensions.  In Sec. \ref{sec:wlp}, we demonstrate that this same regulator dependence arises in the contribution to $H\rightarrow\gamma\gamma$ mediated by SM gauge boson loops.  Then in Sec. \ref{sec:hier} we explore the implications for the hierarchy problem of insisting on obtaining gauge-invariant results for $H\rightarrow\gamma\gamma$ while using a $d=4$ regulator.   We then consider specific solutions to the hierarchy problem and pay particular attention to the case of supersymmetry.  In Sec. \ref{sec:dis} we discuss other possible consequences of our results and conclude.

\section{Fermion and Scalar Loops}
\label{sec:fands}

We now investigate the source of the regulator dependence $H\rightarrow\gamma\gamma$.  In this section, we will consider the contributions to $H\rightarrow\gamma\gamma$ mediated by fermion and scalar loops.  Throughout this paper, we will represent the photon momenta as $q_1$ and $q_2$, with polarization vectors $\varepsilon_1^{*\mu}$ and $\varepsilon_2^{*\nu}$, respectively.  The momentum of the Higgs boson will be denoted $p_H$, and internal loop momenta will be denoted $p$.

\begin{figure}
\epsfig{file=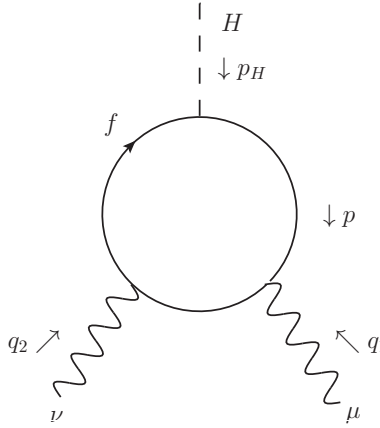,width=2.2in}
\caption{Fermion loop contribution to $H\rightarrow\gamma\gamma$.  Another diagram exists with $\mu\leftrightarrow\nu$, $q_1\leftrightarrow q_2$.}
\label{fig:fdiag}
\end{figure}

We first consider the contribution to $H\rightarrow\gamma\gamma$ from a fermion loop, as shown in Fig. \ref{fig:fdiag}.  In addition to the contribution shown explicitly in Fig. \ref{fig:fdiag}, there is a second diagram with the two photons interchanged, $q_1\leftrightarrow q_2$, $\mu\leftrightarrow \nu$.  The amplitude for the sum of the two fermion diagrams gives
\begin{align}
\label{eq:ferm1}
i {\cal M}^f_{\mu\nu}\varepsilon_1^{*\mu}\varepsilon_2^{*\nu} = \varepsilon_1^{*\mu}\varepsilon_2^{*\nu} \frac{-\lambda_f}{\sqrt{2}}e_f^2\int\frac{d^dp}{(2\pi)^d} \mbox{Tr}\left[\frac{1}{\slashed{p}+\slashed{q}_1+\slashed{q}_2-m_f}\gamma_{\nu}\frac{1}{\slashed{p}+\slashed{q}_1-m_f}\gamma_{\mu}\frac{1}{\slashed{p}-m_f}\right.\nonumber\\\left.+\frac{1}{\slashed{p}+\slashed{q}_1+\slashed{q}_2-m_f}\gamma_{\mu}\frac{1}{\slashed{p}+\slashed{q}_2-m_f}\gamma_{\nu}\frac{1}{\slashed{p}-m_f}\right]
\end{align}
where the two terms in the trace correspond to the two Feynman diagrams.  Here, $e_f$ is the fermion charge, $m_f$ is the fermion mass, and $\lambda_f$ is the Yukawa coupling, which is $\lambda_f=\sqrt{2}m_f/v$, where $v\sim 246$ GeV is the Higgs vacuum expectation value, if the fermion gets its mass entirely from the Higgs.  For the moment, we have kept the dimension of the momentum integral $d$ general.

We now apply the Ward identity to this expression twice\footnote{Although we expect that similar conclusions would be reached if we only applied the Ward identity to a single external photon, applying the Ward identity to both photons makes the relation between gauge invariance in $H\rightarrow\gamma\gamma$ and the hierarchy problem more transparent.}, replacing both photon polarization vectors with their respective momenta, $\varepsilon_1^{*\mu},\varepsilon_2^{*\nu}\rightarrow q_1^{\mu}, q_2^{\nu}$.  For the first term in the trace, we substitute $\slashed{q}_1=(\slashed{p}+\slashed{q}_1-m_f)-(\slashed{p}-m_f)$, $\slashed{q}_2=(\slashed{p}+\slashed{q}_1+\slashed{q}_2-m_f)-(\slashed{p}+\slashed{q}_1-m_f)$ to obtain 
\begin{align}
\label{eq:ferm2}
&\frac{1}{\slashed{p}+\slashed{q}_1+\slashed{q}_2-m_f}\slashed{q}_2\frac{1}{\slashed{p}+\slashed{q}_1-m_f}\slashed{q}_1\frac{1}{\slashed{p}-m_f}\nonumber\\=&\left[\frac{1}{\slashed{p}+\slashed{q}_1+\slashed{q}_2-m_f}\left((\slashed{p}+\slashed{q}_1+\slashed{q}_2-m_f)-(\slashed{p}+\slashed{q}_1-m_f)\right)\frac{1}{\slashed{p}+\slashed{q}_1-m_f}\right.\\& \left.\left((\slashed{p}+\slashed{q}_1-m_f)-(\slashed{p}-m_f)\right)  \frac{1}{\slashed{p}-m_f}\right]\nonumber\\=&\frac{1}{\slashed{p}-m_f}-\frac{1}{\slashed{p}+\slashed{q}_1-m_f}-\frac{1}{\slashed{p}+\slashed{q}_1+\slashed{q}_2-m_f}(\slashed{p}+\slashed{q}_1-m_f)\frac{1}{\slashed{p}-m_f}+\frac{1}{\slashed{p}+\slashed{q}_1+\slashed{q}_2-m_f}.\nonumber
\end{align}
Similarly, subsituting $\slashed{q}_1=(\slashed{p}+\slashed{q}_1+\slashed{q}_2-m_f)-(\slashed{p}+\slashed{q}_2-m_f)$, $\slashed{q}_2=(\slashed{p}+\slashed{q}_2-m_f)-(\slashed{p}-m_f)$, the second term in the trace becomes
\begin{align}
\label{eq:ferm3}
&\frac{1}{\slashed{p}+\slashed{q}_1+\slashed{q}_2-m_f}\slashed{q}_1\frac{1}{\slashed{p}+\slashed{q}_2-m_f}\slashed{q}_2\frac{1}{\slashed{p}-m_f}\nonumber=\\&\frac{1}{\slashed{p}-m_f}-\frac{1}{\slashed{p}+\slashed{q}_2-m_f}-\frac{1}{\slashed{p}+\slashed{q}_1+\slashed{q}_2-m_f}(\slashed{p}+\slashed{q}_2-m_f)\frac{1}{\slashed{p}-m_f}+\frac{1}{\slashed{p}+\slashed{q}_1+\slashed{q}_2-m_f}.
\end{align}
Summing these expressions and substituting them back into $i{\cal M}^f_{\mu\nu}q_1^{\mu}q_2^{\nu}$, we obtain
\begin{align}
\label{eq:fermdshift}
i{\cal M}^f_{\mu\nu}q_1^{\mu}q_2^{\nu}=\frac{-\lambda_f}{\sqrt{2}}e_f^2\int \frac{d^dp}{(2\pi)^d}\mbox{Tr}&\left[\frac{1}{\slashed{p}-m_f} - \frac{1}{\slashed{p}+\slashed{q}_1-m_f}\right.\nonumber\\ &\left.-\frac{1}{\slashed{p}+\slashed{q}_2-m_f} + \frac{1}{\slashed{p}+\slashed{q}_1+\slashed{q}_2-m_f}\right].
\end{align}

\begin{figure}
\epsfig{file=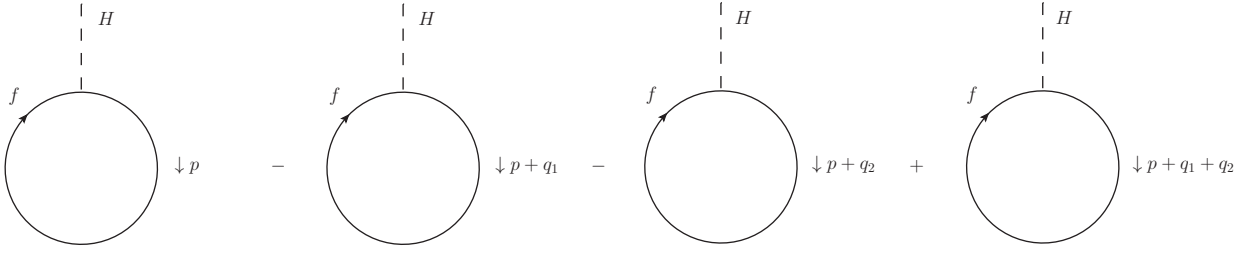,width=6.5in}
\caption{The diagrammatic representation, up to a factor of $e_f^2$, of the expression obtained by applying the Ward Identity to the fermionic loop contribution to $H\rightarrow\gamma\gamma$.}
\label{fig:ftdiag}
\end{figure}

We now compare the expression in Eq. (\ref{eq:fermdshift}) to that obtained from Higgs tadpole diagrams, such as those shown in Fig. \ref{fig:ftdiag}.  The amplitude for a single tadpole diagram with fermion loop momentum $p$ is
\begin{equation}
\label{eq:ftad}
i {\cal M}^f_{tadpole} = \frac{-\lambda_f}{\sqrt{2}} \int \frac{d^dp}{(2\pi)^d}\mbox{Tr}\left[\frac{1}{\slashed{p}-m_f}\right].
\end{equation}
We thus notice that the result of applying the Ward identity to both external photons gives precisely $e_f^2$ times that obtained from the combination of tadpole diagrams shown in Fig. \ref{fig:ftdiag}, which differ from each other only in the definition of their loop momenta.  We note that these tadpole diagrams are quadratically-divergent, as the trace in Eq. (\ref{eq:ftad}) equals $4m_f/(p^2-m_f^2)$.

This relation between electromagnetic gauge invariance in $H\rightarrow\gamma\gamma$ and changes in Higgs tadpole diagrams under shifts of loop momenta gives us the first indication of the cause of the regulator-dependence in $H\rightarrow\gamma\gamma$; dimensional regularization is shift-invariant, so the difference shown in Fig. \ref{fig:ftdiag} gives zero.  However, this does not necessarily hold for $d=4$.  To see this more explicitly, we return to Eq. (\ref{eq:ferm1}).  We note that all of the terms in this expression are at most logarithmically divergent.  (Although the integral may initially appear to be linearly-divergent, one may note that nonzero terms in the trace must contain an even number of gamma matrices; this implies that all terms will contain at least one factor of $m_f$ and lowers the overall divergence to at most logarithmic order.)  Also, as we are only interested in terms which differ in going from $d=4$ to $d=4-\epsilon$, we need not consider finite terms.  However, the divergence structure in $H\rightarrow\gamma\gamma$ is independent of external momenta; thus, to explore the regulator dependence of Eq. (\ref{eq:ferm1}), we will greatly simplify the calculation of all amplitudes by setting $q_1, q_2=0$. 

Evaluating Eq. (\ref{eq:ferm1}) for $q_1, q_2=0$, we obtain
\begin{align}
\label{eq:ferm4}
\left.i {\cal M}^f_{\mu\nu}\right|_{q_{1,2}=0}\varepsilon_1^{*\mu}\varepsilon_2^{*\nu} = \varepsilon_1^{*\mu}\varepsilon_2^{*\nu} \frac{-\lambda_f}{\sqrt{2}}e_f^2\int\frac{d^dp}{(2\pi)^d} \mbox{Tr}\left[\frac{1}{\slashed{p}-m_f}\gamma_{\nu}\frac{1}{\slashed{p}-m_f}\gamma_{\mu}\frac{1}{\slashed{p}-m_f} + \mu \leftrightarrow \nu\right].
\end{align}
Considering just the momentum integral, we obtain
\begin{align}
\label{eq:ferm5}
&\int\frac{d^dp}{(2\pi)^d} \mbox{Tr}\left[\frac{1}{\slashed{p}-m_f}\gamma_{\nu}\frac{1}{\slashed{p}-m_f}\gamma_{\mu}\frac{1}{\slashed{p}-m_f} + (\mu \leftrightarrow \nu)\right]\nonumber\\=&\int\frac{d^dp}{(2\pi)^d} \frac{1}{(p^2-m_f^2)^3}\mbox{Tr}\left[(\slashed{p}+m_f)\gamma_{\nu}(\slashed{p}+m_f)\gamma_{\mu}(\slashed{p}+m_f) + (\mu \leftrightarrow \nu)\right]\\=&\int\frac{d^dp}{(2\pi)^d} \frac{1}{(p^2-m_f^2)^3}\mbox{Tr}\left[m_f(2\slashed{p}\gamma_{\nu}\slashed{p}\gamma_{\mu} + \slashed{p}\gamma_{\nu}\gamma_{\mu}\slashed{p}) +m_f^3(\gamma_{\nu}\gamma_{\mu})+ (\mu \leftrightarrow \nu)\right].\nonumber
\end{align}
Evaluation of the trace is independent of $d$, and we obtain
\begin{equation}
\label{eq:ferm6}
(8m_f)\int\frac{d^dp}{(2\pi)^d} \frac{4p_{\mu}p_{\nu}-g_{\mu\nu}(p^2-m_f^2)}{(p^2-m_f^2)^3}.
\end{equation}
The integral in Eq. (\ref{eq:ferm6}) is central to this paper.  If evaluated for $d=4$, one can do the substitution $4p_{\mu}p_{\nu}\rightarrow p^2g_{\mu\nu}$ and obtain
\begin{equation}
\label{eq:amb4d}
\int\frac{d^4p}{(2\pi)^4} \frac{4p_{\mu}p_{\nu}-g_{\mu\nu}(p^2-m_f^2)}{(p^2-m_f^2)^3}=\int\frac{d^4p}{(2\pi)^4} \frac{g_{\mu\nu}m_f^2}{(p^2-m_f^2)^3}=\frac{i}{(4\pi)^2}\left(-\frac{g_{\mu\nu}}{2}\right)\ne 0.
\end{equation}
On the other hand, if one uses $d=4-\epsilon$, $4p_{\mu}p_{\nu}\rightarrow 4/(4-\epsilon)p^2g_{\mu\nu}$, and we get
\begin{align}
\label{eq:ambdr1}
\int\frac{d^{4-\epsilon}p}{(2\pi)^{4-\epsilon}} \frac{4p_{\mu}p_{\nu}-g_{\mu\nu}(p^2-m_f^2)}{(p^2-m_f^2)^3}=\int\frac{d^{4-\epsilon}p}{(2\pi)^{4-\epsilon}} \frac{g_{\mu\nu}(\frac{\epsilon}{4} p^2+m_f^2)}{(p^2-m_f^2)^3}.
\end{align}
Using 
\begin{align}
\label{eq:ambdr2}
\int\frac{d^{4-\epsilon}p}{(2\pi)^{4-\epsilon}} \frac{p^2}{(p^2-m_f^2)^3}=\frac{i}{(4\pi)^2}\left(\frac{2}{\epsilon}+\mbox{finite}\right),
\end{align}
we find
\begin{align}
\label{eq:ambdr}
\int\frac{d^{4-\epsilon}p}{(2\pi)^{4-\epsilon}} \frac{4p_{\mu}p_{\nu}-g_{\mu\nu}(p^2-m_f^2)}{(p^2-m_f^2)^3}=0.
\end{align}
We thus see that the evaluation of the integral in Eq. (\ref{eq:ferm6}) differs in going from $d=4$ to $d=4-\epsilon$.  We now relate this back to the behavior of Higgs tadpole diagrams under shifts of loop momenta.  As we show in Appendix \ref{appendix:appa}, 
\begin{align}
\label{eq:quadshift}
&\int\frac{d^dp}{(2\pi)^d}\left(\frac{1}{p^2-m^2}-\frac{1}{(p+q_1)^2-m^2}-\frac{1}{(p+q_2)^2-m^2}+\frac{1}{(p+q_1+q_2)^2-m^2}\right)\nonumber\\&=(2)q_1^{\mu}q_2^{\nu}\int\frac{d^dp}{(2\pi)^d}\frac{4p_{\mu}p_{\nu}-g_{\mu\nu}(p^2-m^2)}{(p^2-m^2)^3}.
\end{align}
Thus, comparing the result in Eq. (\ref{eq:quadshift}) to to those in Eqs. (\ref{eq:amb4d}) and (\ref{eq:ambdr}), we see explicitly that the combination of tadpole diagrams shown in Fig. \ref{fig:ftdiag} gives zero when evaluated in dimensional regularization, but gives a nonzero result when evaluated for $d=4$.  

To make the source of the difference in behavior between $d=4$ and $d=4-\epsilon$ more transparent, let us consider just the first two terms in the difference in Eq. (\ref{eq:quadshift}),
\begin{equation}
\label{eq:singshift}
\int\frac{d^dp}{(2\pi)^d}\left(\frac{1}{p^2-m^2}-\frac{1}{(p+q)^2-m^2}\right)=\int\frac{d^dp}{(2\pi)^d}\frac{2p\cdot q +q^2}{(p^2-m^2)((p+q)^2-m^2)}.
\end{equation}
Thus, the combination of the four tadpole terms given in Eq. (\ref{eq:quadshift}) can be written as a difference between two terms
\begin{align}
\label{eq:lin}
\int\frac{d^dp}{(2\pi)^d}\left(\frac{2p\cdot q_1 +q_1^2}{(p^2-m^2)((p+q_1)^2-m^2)} - \frac{2(p+q_2)\cdot q_1 +q_1^2}{((p+q_2)^2-m^2)((p+q_1+q_2)^2-m^2)}\right)
\end{align}
which themselves differ from each other by only a momentum shift, $p\rightarrow p+q_2$.  (Explicitly, the difference in Eq. (\ref{eq:quadshift}) is obtained if we take the change in the tadpole diagram when the loop momentum is shifted by $q_1$ and then compute the change in the resulting expression under a loop momentum shift of $q_2$.  We will loosely refer to this as performing successive momentum shifts of $q_1$ and $q_2$.)  We now consider a point shown in \cite{Elias:1982sq}, namely that linearly-divergent integrals are not invariant under loop momentum shifts, but that less-than-linearly-divergent integrals are.  Thus when Eq. (\ref{eq:lin}) is evaluated with $d=4$, it is the change under a loop momentum shift of a linearly-divergent integral.  However, when dimensional regularization is used, setting $d=4-\epsilon$ renders these terms less-than-linearly divergent, and thus shift-invariant; thus, for $d=4-\epsilon$, the expression in Eq. (\ref{eq:lin}) vanishes.  And, thus, we see why gauge-invariance is violated in going from $d=4-\epsilon$ to $d=4$:  although the Feynman diagrams for $H\rightarrow\gamma\gamma$ are themselves invariant under shifts in the loop momenta, applying the QED Ward identity to $H\rightarrow\gamma\gamma$ yields an expression which is equal to the change in a divergent integral under a shift in loop momentum.  For $d=4$, this integral is sufficiently divergent that shifts of loop momenta change the result and thus spoil gauge invariance; for $d=4-\epsilon$, the integral is shift-invariant, and no such loss of gauge invariance occurs.  

\begin{figure}
\epsfig{file=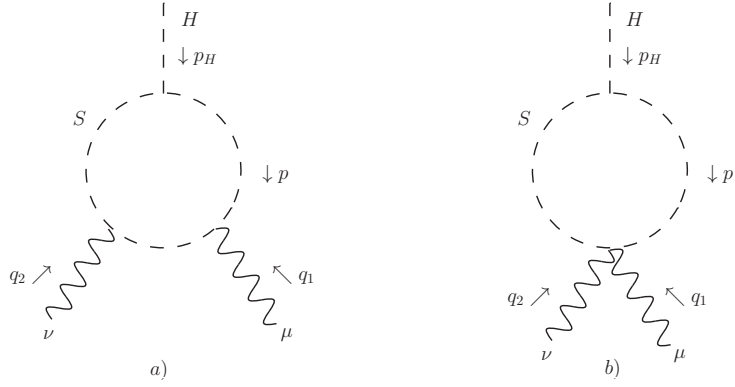,width=4in}
\caption{Scalar loop contributions to $H\rightarrow\gamma\gamma$.  Another diagram exists identical to a) but with $\mu\leftrightarrow\nu$, $q_1\leftrightarrow q_2$.}
\label{fig:sdiag}
\end{figure}

We now repeat the above analysis for a scalar loop, showing an analogous relation between the application of the Ward identity to $H\rightarrow\gamma\gamma$ and the behavior of Higgs tadpole diagrams under shifts of loop momenta.  The diagrams contributing to $H\rightarrow\gamma\gamma$ are shown in Fig. \ref{fig:sdiag}; like in the fermion loop case, there is also a diagram (not shown) which is identical to that in Fig. \ref{fig:sdiag} a) but with the two external photons interchanged.  Writing the four-scalar term in the Lagrangian\footnote{We choose this normalization for $\lambda_S$ as it gives the same Feynman rule for the triple-scalar vertex as that for the Higgs-Goldstone-Goldstone vertex derived from ${\cal L}=-\lambda \phi^{\dagger}\phi\phi^{\dagger}\phi$.} as ${\cal L}_{SSHH}=-2\lambda_SS^+S^-\phi^{\dagger}\phi$, where $\phi$ is the SM Higgs doublet, the Feynman rule for the $HS^+S^-$ vertex is $-2i\lambda_S v$, and the sum of the amplitudes gives
\begin{align}
\label{eq:scal1}
i {\cal M}^S_{\mu\nu}\varepsilon_1^{*\mu}\varepsilon_2^{*\nu} = &\varepsilon_1^{*\mu}\varepsilon_2^{*\nu} 2 \lambda_S v e_S^2\int\frac{d^dp}{(2\pi)^d} \frac{1}{p^2-m_S^2} \frac{1}{(p+q_1+q_2)^2-m_S^2}\nonumber\\ &\times\left[ \frac{(2p+q_1)_{\mu}(2p+2q_1+q_2)_{\nu}}{(p+q_1)^2-m_S^2} + \frac{(2p+q_2)_{\nu}(2p+2q_2+q_1)_{\mu}}{(p+q_2)^2-m_S^2}-2g_{\mu\nu}\right]. 
\end{align} 

We first note that, like the case of the fermion loop, terms in Eq. (\ref{eq:scal1}) are at most logarithmically divergent, and this divergence structure is independent of the external momenta.  Thus, if we are interested in only those terms which can depend on the regulator, we can set $q_1=q_2=0$.  Doing so yields
\begin{align}
\label{eq:scal2}
\left.i {\cal M}^S_{\mu\nu}\right|_{q_{1,2}=0}\varepsilon_1^{*\mu}\varepsilon_2^{*\nu} = \varepsilon_1^{*\mu}\varepsilon_2^{*\nu} 4 \lambda_S v e_S^2\int\frac{d^dp}{(2\pi)^d} \frac{4p_{\mu}p_{\nu}-g_{\mu\nu}(p^2-m_S^2)}{(p^2-m_S^2)^3}.
\end{align}
Thus, we see that the same regulator-dependent integral occurs as in the fermion case.

\begin{figure}
\epsfig{file=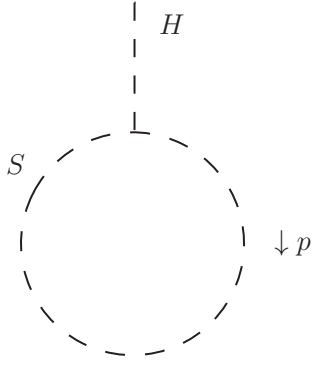,width=1.9in}
\caption{Scalar loop contribution to the Higgs tadpole.}
\label{fig:stdiag}
\end{figure}

Returning now to the case of general $q_1$, $q_2$, we apply the Ward identity to both of the external photons, replacing their polarization vectors in Eq. (\ref{eq:scal1}) with their respective momenta.  
\begin{align}
\label{eq:scal3}
i {\cal M}^S_{\mu\nu}q_1^{\mu}q_2^{\nu} = & 2 \lambda_S v e_S^2\int\frac{d^dp}{(2\pi)^d} \frac{1}{p^2-m_S^2} \frac{1}{(p+q_1+q_2)^2-m_S^2}\nonumber\\ \times&\left[ \frac{((p+q_1)^2-p^2)((p+q_1+q_2)^2-(p+q_1)^2)}{(p+q_1)^2-m_S^2}\right.\nonumber\\& \left.+ \frac{((p+q_2)^2-p^2)((p+q_1+q_2)^2-(p+q_2)^2)}{(p+q_2)^2-m_S^2}-2q_1\cdot q_2\right]. 
\end{align} 
After straightforward simplification, this becomes
\begin{align}
\label{eq:scal4}
&i {\cal M}^S_{\mu\nu}q_1^{\mu}q_2^{\nu} =  2 \lambda_S v e_S^2\nonumber\\&\times\int\frac{d^dp}{(2\pi)^d} \left[\frac{1}{p^2-m_S^2} -\frac{1}{(p+q_1)^2-m_S^2} - \frac{1}{(p+q_2)^2-m_S^2} + \frac{1}{(p+q_1+q_2)^2-m_S^2}\right].
\end{align}
We can now compare this to the amplitude for the scalar loop contribution to a Higgs tadpole, shown in Fig. \ref{fig:stdiag},
\begin{equation}
\label{eq:scalt}
i{\cal M}^S_{tadpole} =2\lambda_S v \int\frac{d^dp}{(2\pi)^d} \frac{1}{p^2-m_S^2,}
\end{equation}
and we see that applying the Ward identity to both photons in the scalar loop contribution to $H\rightarrow\gamma\gamma$ yields behavior essentially identical to that of the fermion loop contribution.  Specifically, applying the Ward identity to both photons yields a regulator-dependent expression which, up to a factor of the square of the loop particle's charge, is equal to the change of the corresponding Higgs tadpole diagram subjected to two successive loop momentum shifts of $q_1$ and $q_2$.

We will now extend this analysis to the case of the SM $W^{\pm}$ loop.

\section{$W$ Loop}
\label{sec:wlp}
\begin{figure}
\epsfig{file=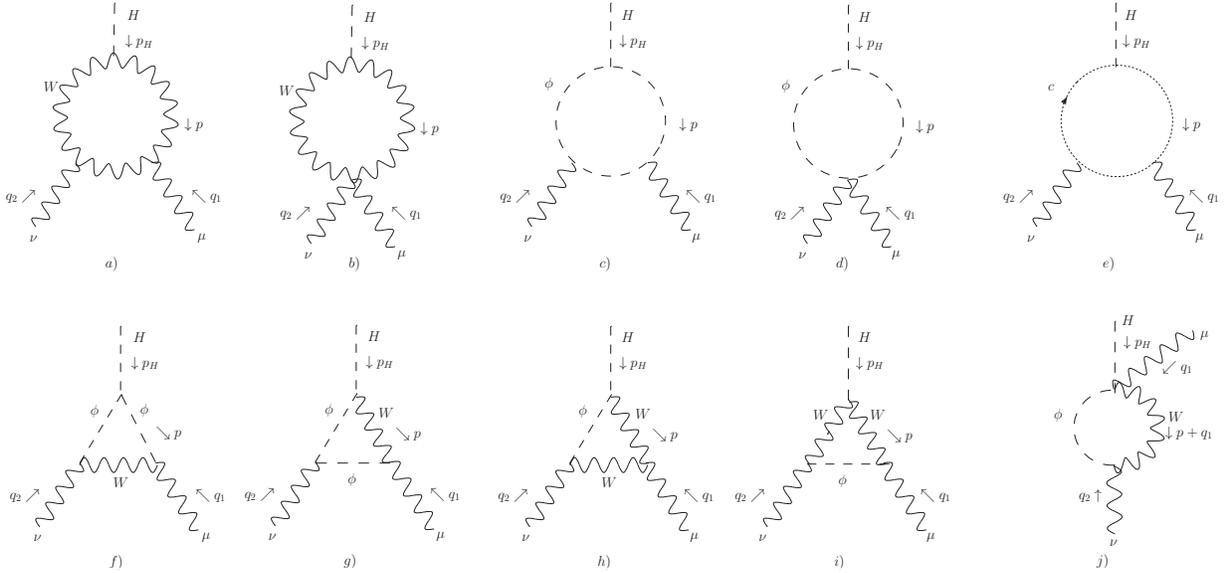,width=6.5in}
\caption{Gauge boson loop contributions to $H\rightarrow\gamma\gamma$.  Diagrams also exist identical to a), c), e), f), and i) but with $\mu\leftrightarrow\nu$, $q_1\leftrightarrow q_2$.  Diagram e) should be taken as a sum over both charged ghost fields.  Diagrams identical to g), h), and j) exist but with $\mu\leftrightarrow\nu$, $q_1\leftrightarrow q_2$ and/or the internal lines attached to $H$ interchanged.}
\label{fig:wdiag}
\end{figure}

We now wish to show that the behavior observed for the fermionic and scalar contributions to $H\rightarrow\gamma\gamma$ also occurs for the contribution from the SM $W^{\pm}$ loop.  We work in general $R_{\xi}$ gauge, and thus must include diagrams which contain charged Goldstone bosons and charged ghost fields.  These diagrams are shown in Fig. \ref{fig:wdiag}; momenta are defined as in the case of the fermion and scalar loops.

\begin{figure}
\epsfig{file=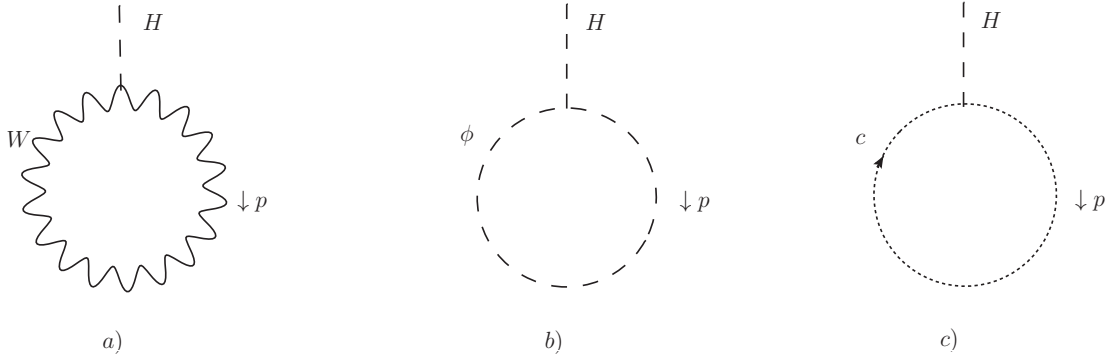,width=6.0in}
\caption{Higgs tadpole contributions from loops of a) the $W^{\pm}$ boson, b) chaeged Goldstone bosons, and c) charged ghosts in renormalizeable gauge.}
\label{fig:wtdiag}
\end{figure}

Because we work in $R_{\xi}$ gauge but do not take the unitary $\xi\rightarrow \infty$ limit, the diagrams in Fig. \ref{fig:wdiag} contain only logarithmically divergent and finite terms.  As in the fermion and scalar loop cases, all divergent, and therefore possibly regulator-dependent terms, are independent of the external momenta.  Thus, we will forgo the complete calculation of $H\rightarrow\gamma\gamma$ and instead calculate only the amplitudes setting $p_H=q_1=q_2=0$; we then examine the difference obtained in going from $d=4-\epsilon$ to $d=4$.  As the change in the amplitude in going from $d=4$ to $d=4-\epsilon$ is independent of the external momenta, we can then apply this result to the physical case where all external particles are on-shell, $p_H^2=m_H^2$, $q_1^2=q_2^2=0$.  We take the physical case to obey the Ward identity when evaluated with dimensional regularization, $d=4-\epsilon$; thus, any terms that violate the Ward identity in $d=4$ will be contained in the regulator-dependent terms in the amplitude.  We then show that the difference between the $d=4-\epsilon$ and $d=4$ cases is reflected in the behavior of the tadpoles in Fig. \ref{fig:wtdiag} under successive momentum shifts of $q_1$ and $q_2$.

We present our results for ${\cal M}^i_{\mu\nu}$ for each of these diagrams below.  A few notes are in order about these various terms.  First, additional diagrams exist which can be obtained from those in Fig. \ref{fig:wdiag} by exchange of external photons or by exchange of internal lines.  Second, we will explicitly write the momentum integrals in $d=4-\epsilon$ dimensions; for the case of $d=4$, $\epsilon$ should be set to $0$.  Lastly, we note that Eqs. (\ref{eq:wdiaga}-\ref{eq:appbsum}) are all evaluated for $q_1=q_2=p_H=0$, even though, for simplicity, we have not labelled them as such.

We begin with the diagrams which contain only $W$ bosons, diagrams a) and b).  In the case of diagram a), we must also include the contribution from the diagram with the external photons interchanged, $q_1\leftrightarrow q_2$, $\mu\leftrightarrow\nu$.  The sum of these two contributions gives 
\begin{align}
i{\cal M}^a_{\mu\nu}\varepsilon^{*\mu}_1\varepsilon^{*\nu}_2=\varepsilon^{*\mu}_1\varepsilon^{*\nu}_2(e^2 g M_W)\int \frac{d^dp}{(2\pi)^d}  \frac{2}{(p^2-M_W^2)^3}\left[\vphantom{\frac{(1-\xi)}{(p^2-\xi M_W^2)}} \left(2p^2 g_{\mu\nu} + (10-4\epsilon) p_{\mu}p_{\nu}\right)\right.\nonumber\\ - \left.\frac{(1-\xi)}{(p^2-\xi M_W^2)}(3p^4 g_{\mu\nu} -3 p^2 p_{\mu}p_{\nu})  + \frac{(1-\xi)^2}{(p^2-\xi M_W^2)^2}(p^6 g_{\mu\nu} -p^4 p_{\mu}p_{\nu})\right],
\label{eq:wdiaga}
\end{align}
and 
\begin{eqnarray}
&&i{\cal M}^b_{\mu\nu}\varepsilon^{*\mu}_1\varepsilon^{*\nu}_2=\varepsilon^{*\mu}_1\varepsilon^{*\nu}_2(e^2 g M_W)\int \frac{d^dp}{(2\pi)^d}  \frac{-1}{(p^2-M_W^2)^2}\times\nonumber\\ &&\left[(6-2\epsilon)g_{\mu\nu}-2\frac{(1-\xi)}{(p^2-\xi M_W^2)}(2 p^2 g_{\mu\nu} - 2p_{\mu}p_{\nu}) +\frac{(1-\xi)^2}{(p^2-\xi M_W^2)^2}(2p^4g_{\mu\nu}-2p^2p_{\mu}p_{\nu})\right] .
\label{eq:wdiagb}
\end{eqnarray}
These expressions contain an explicit dependence on $\epsilon$, which results from the contraction of a metric tensor, $g^{\mu}_{\mu}=d$.  Note that these terms will cancel in the sum.

We now consider the diagrams which contain only Goldstone bosons on the internal lines, c) and d).  This follows exactly as the scalar case discussed in Sec. \ref{sec:fands} with $\lambda_S=m_H^2/2v^2$, and we obtain, for the two contributions represented by c) 
\begin{equation}
i{\cal M}^c_{\mu\nu}\varepsilon^{*\mu}_1\varepsilon^{*\nu}_2=\varepsilon^{*\mu}_1\varepsilon^{*\nu}_2\left(\frac{e^2 m_H^2 g}{M_W}\right)\int \frac{d^dp}{(2\pi)^d} 4 \frac{p_{\mu}p_{\nu}}{(p^2-\xi M_W^2)^3};
\label{eq:wdiagc}
\end{equation}
while, for d), we obtain
\begin{equation}
i{\cal M}^d_{\mu\nu}\varepsilon^{*\mu}_1\varepsilon^{*\nu}_2=\varepsilon^{*\mu}_1\varepsilon^{*\nu}_2\left(\frac{e^2 m_H^2 g}{M_W}\right)\int \frac{d^dp}{(2\pi)^d} \left[-\frac{g_{\mu\nu}}{(p^2-\xi M_W^2)^2}\right].
\label{eq:wdiagd}
\end{equation}
We note that these contributions sum to
\begin{equation}
i{\cal M}^{goldstone}_{\mu\nu}\varepsilon^{*\mu}_1\varepsilon^{*\nu}_2=\varepsilon^{*\mu}_1\varepsilon^{*\nu}_2\left(\frac{e^2 m_H^2 g}{M_W}\right)\int \frac{d^dp}{(2\pi)^d} \left[\frac{4p_{\mu}p_{\nu}-g_{\mu\nu}(p^2-\xi M_W^2)}{(p^2-\xi M_W^2)^3}\right],
\label{eq:wdiaggoldsum}
\end{equation}
which is in agreement with our results of Section \ref{sec:fands}.

Next, we consider the ghost loops, represented by diagram e).  There are actually four diagrams, obtained by exchanging the external photon lines and summing over the postively- and negatively-charged ghost fields, $c^{\pm}$.  For their sum, we obtain
\begin{equation}
i{\cal M}^e_{\mu\nu}\varepsilon^{*\mu}_1\varepsilon^{*\nu}_2=\varepsilon^{*\mu}_1\varepsilon^{*\nu}_2(e^2 g M_W)\int \frac{d^dp}{(2\pi)^d}  (-2\xi)\frac{p_{\mu}p_{\nu}}{(p^2-\xi M_W^2)^3}.
\label{eq:wdiage}
\end{equation}

We now consider the remaining diagrams.  Diagram f), along with a second diagram obtained by exchanging the external photon lines, is finite, and, thus regulator-independent; choosing to include or not include this contribution makes no difference in the final result, and, for simplicity, we do not include it here.  Diagram g) represents four contributions, obtained by exchanging either the external photon lines or the internal lines attached to the Higgs.  They sum to
\begin{align}
i{\cal M}^g_{\mu\nu}\varepsilon^{*\mu}_1\varepsilon^{*\nu}_2=&\varepsilon^{*\mu}_1\varepsilon^{*\nu}_2(e^2 g M_W)\nonumber\\&\times\int \frac{d^dp}{(2\pi)^d}  \frac{4}{(p^2-\xi M_W^2)^2} \frac{1}{(p^2- M_W^2)}  \left[ p_{\mu}p_{\nu} -(1-\xi)\frac{p^2 p_{\mu}p_{\nu}}{(p^2-\xi M_W^2)}\right].
\label{eq:wdiagg}
\end{align}
Diagram h) similarly represents four contributions, which sum to
\begin{equation}
i{\cal M}^h_{\mu\nu}\varepsilon^{*\mu}_1\varepsilon^{*\nu}_2=\varepsilon^{*\mu}_1\varepsilon^{*\nu}_2(e^2 g M_W)\int \frac{d^dp}{(2\pi)^d} (-2\xi) \frac{1}{(p^2-\xi M_W^2)^2} \frac{1}{(p^2- M_W^2)} \left[p^2 g_{\mu\nu} -p_{\mu}p_{\nu} \right].
\label{eq:wdiagh}
\end{equation}
The two contributions represented by diagram i) are finite and, thus, regulator-independent; like the case of diagram f) above, as long as one considers the difference obtained in going from $d=4-\epsilon$ to $d=4$, the choice to include or not include diagram i) makes no difference in the final result.  However, including this diagram simplifies the calculation, so we retain it here; we obtain 
\begin{align}
i{\cal M}^i_{\mu\nu}\varepsilon^{*\mu}_1\varepsilon^{*\nu}_2=&\varepsilon^{*\mu}_1\varepsilon^{*\nu}_2(e^2 g M_W)\int \frac{d^dp}{(2\pi)^d} (-2M_W^2) \frac{1}{(p^2-\xi M_W^2)} \frac{1}{(p^2- M_W^2)^2}\nonumber \\&\times\left[g_{\mu\nu} -(1-\xi)\frac{2p_{\mu}p_{\nu}}{(p^2-\xi M_W^2)} + (1-\xi)^2 \frac{p^2p_{\mu}p_{\nu}}{(p^2-\xi M_W^2)^2}\right].
\label{eq:wdiagi}
\end{align}
Lastly, for the four diagrams represented by j) (obtained by exchanging the external photons and/or the internal lines), we obtain
\begin{align}
i{\cal M}^j_{\mu\nu}\varepsilon^{*\mu}_1\varepsilon^{*\nu}_2=&\varepsilon^{*\mu}_1\varepsilon^{*\nu}_2(e^2 g M_W)\nonumber\\&\times\int \frac{d^dp}{(2\pi)^d}  \frac{-2}{(p^2-\xi M_W^2)} \frac{1}{(p^2- M_W^2)} \left[g_{\mu\nu} -(1-\xi) \frac{p_{\mu}p_{\nu}}{(p^2-\xi M_W^2)} \right].
\label{eq:wdiagj}
\end{align}
We have checked that the expressions for diagrams a), b,) and e)-j), for $\xi=1$, reduce to the divergent terms listed in \cite{Ellis:1975ap} and that the sum of the expressions for diagrams c) and d) agrees with that given in \cite{Marciano:2011gm}.  We now examine these results more closely and compare them to those of the tadpole diagrams shown in Fig. \ref{fig:wtdiag}.  The expressions for these three tadpole diagrams are
\begin{align}
i{\cal M}_{tadpole}^{W}&=gM_W\int \frac{d^dp}{(2\pi)^d}\frac{1}{p^2-M_W^2}\left((4-\epsilon)-\frac{p^2(1-\xi)}{p^2-\xi M_W^2}\right),\nonumber\\
i{\cal M}_{tadpole}^{\phi}&=\left(\frac{g m_H^2}{M_W}\right)\int \frac{d^dp}{(2\pi)^d} \left(\frac{1}{2}\right)\frac{1}{(p^2-\xi M_W^2)},\\
i{\cal M}_{tadpole}^{c}&=gM_W\int \frac{d^dp}{(2\pi)^d}\frac{-\xi}{p^2-\xi M_W^2}\nonumber.
\end{align}
First, we examine the diagrams containing only Goldstone particles on the internal legs, Figs. \ref{fig:wdiag} c) and d), and \ref{fig:wtdiag} b).  From our discussion of the scalar loop case in Sec. \ref{sec:fands}, we can see immediately that the Ward Identity, applied to Figs. \ref{fig:wdiag} c) and d), gives $e^2$ times a double-shift of \ref{fig:wtdiag} b).  Thus, we will concentrate on the remaining diagrams below.

We note that the $W$ and ghost tadpoles sum\footnote{One may worry that this sum is not well-defined due to momentum-routing amibiguities in the quadratically-divergent tadpole diagrams.  Although these ambiguities do exist, we will be concerned with differences in tadpole diagrams similar to those shown in Fig. \ref{fig:ftdiag}; when these differences are taken, the resulting expressions contain terms which are at most logarithmically divergent, and thus invariant under momentum shifts.  Therefore any momentum-routing ambiguities present in Eq. (\ref{eq:wtdiagwc}) do not affect our results.} to a gauge-invariant expression,
\begin{equation}
i{\cal M}_{tadpole}^{W+c}=gM_W\int \frac{d^dp}{(2\pi)^d}  \frac{(3-\epsilon)}{(p^2-M_W^2)}.
\label{eq:wtdiagwc}
\end{equation}
We show in Appendix \ref{appendix:appb} that the sum of diagrams a), b), e)-j) evaluated for $q_1=q_2=p_H=0$ is
\begin{equation}
\label{eq:appbsum}
i{\cal M}^{a,b,e-j}_{\mu\nu}\varepsilon^{*\mu}_1\varepsilon^{*\nu}_2=\varepsilon^{*\mu}_1\varepsilon^{*\nu}_2(e^2 g M_W)\int \frac{d^dp}{(2\pi)^d} (6-2\epsilon)\frac{4p_{\mu}p_{\nu}-g_{\mu\nu}(p^2-M_W^2)}{(p^2-M_W^2)^3} + \mbox{ finite},
\end{equation}
where the ``finite'' term is nonzero and arises from diagram f).  We note that the explicit $-2\epsilon$ term in (\ref{eq:appbsum}) is multiplied by an finite expression, and therefore does not contribute as $\epsilon\rightarrow 0$.  We retain it here for comparison with Eq. (\ref{eq:wtdiagwc}), but will drop it in future expressions. 

We now examine the regulator dependence of Eq. (\ref{eq:appbsum}).  First, we note that the finite terms in Eq. (\ref{eq:appbsum}) do not change in going from $d=4$ to $d=4-\epsilon$.  Second, although we have evaluated this expression for $q_1=q_2=p_H=0$, the expression for $i{\cal M}^{a,b,e-j}_{\mu\nu}\varepsilon^{*\mu}_1\varepsilon^{*\nu}_2$ evaluated for on-shell external states differs from this one only by (complicated) finite terms which, being finite, are also regulator-independent.  Therefore, if we apply the Ward identity to both photons in this amplitude, we obtain
\begin{equation}
\label{eq:appbsumwi1}
\left.i{\cal M}^{a,b,e-j}_{\mu\nu}q_1^{\mu}q_2^{\nu}\right|_{on-shell}=q_1^{\mu}q_2^{\nu}(e^2 g M_W)\int \frac{d^dp}{(2\pi)^d} (6)\frac{4p_{\mu}p_{\nu}-g_{\mu\nu}(p^2-M_W^2)}{(p^2-M_W^2)^3} + \mbox{ finite}.
\end{equation}
Comparing with Eq. (\ref{eq:quadshift}), we see that the only regulator-dependent terms that arise when the Ward identity is applied to both photons is equal to $e^2$ times the difference of the expression in Eq. (\ref{eq:wtdiagwc}) under successive momentum shifts of $q_1$ and $q_2$.  Taking the on-shell amplitude to satisfy the Ward identity when evaluated in $d=4-\epsilon$ and again noting that the integral in Eq. (\ref{eq:ambdr}) is zero when evaluated using dimensional regualrization, we conclude that the finite terms in Eq. (\ref{eq:appbsumwi1}) vanish.  Thus, for $d=4$,
\begin{equation}
\label{eq:appbsumwi2}
\left.i{\cal M}^{a,b,e-j}_{\mu\nu}q_1^{\mu}q_2^{\nu}\right|_{on-shell}=q_1^{\mu}q_2^{\nu}(e^2 g M_W)\int \frac{d^4p}{(2\pi)^4} (6)\frac{4p_{\mu}p_{\nu}-g_{\mu\nu}(p^2-M_W^2)}{(p^2-M_W^2)^3}\quad (d=4),
\end{equation}
which is precisely $e^2$ times the expression for the change in the Higgs tadpole in Eq. (\ref{eq:wtdiagwc}) under successive momentum shifts for $d=4$.

\section{Consquences for Solutions to the Hierarchy Problem}
\label{sec:hier}

We now wish to explore some possible consequences that the above results have for the hierarchy problem.  In order to do this, we must clarify some assumptions that will be relevant for what follows.  For this work, we take the hierarchy problem to be the sensitivity, for certain regulators, of the Higgs mass-squared to quadratically-divergent radiative corrections.  We take these quadratic divergences to be physical and do not use a regulator, such as dimensional regularization, in which both these quadratic divergences and the gauge-invariance-violating terms in $H\rightarrow\gamma\gamma$ do not appear.  We do not consider the sensitivity of $m_H^2$ to finite terms arising from high-scale physics; our results are not relevant for such contributions.  Although the quadratic divergences in the Higgs self-energy could be cancelled within the SM by assuming that parameters of the SM Lagrangian are extremely fine-tuned, we assume that these quadratic divergences will instead be cancelled off by physics beyond the SM, and that this new physics takes the form of new loops of scalar or fermionic particles.\footnote{We note that it is possible that these assumptions are not fulfilled in nature.  For example, one may argue that these quadratic divergences are unphysical and should simply be cancelled by counterterms similarly to the treatment of logarithmic divergences.  Or, alternatively, perhaps these quadratic divergences are physical but cancelled by more exotic physics.  Another possibility is that a true cutoff exists, such as in the case of a composite Higgs.  Our analysis is not applicable to these scenarios.}  For simplicity, we assume that the gauge group of all new physics is that of the SM; we do not consider new gauge boson loops.  

We now investigate the possible relevance of the $d=4$ calculation of $H\rightarrow\gamma\gamma$ to the cancellation of quadratic divergences in the Higgs self-energy.  Here we take the hypothesis that the $d=4$ calculation is valid, but that the gauge-invariance-violating terms derived above cancel when all contributions are included.  From our results above, we see that this can be achieved if the sum of the amplitudes of Higgs tadpole diagrams, weighted by the square of the loop charge, equals zero,
\begin{equation}
\label{eq:grandrelation}
e^23gM_W+\frac{e^2gm_H^2}{2M_W} +\displaystyle\sum_{\rm scalars} e_s^2 (2\lambda_S v)-\displaystyle\sum_{\rm fermions}e_f^2 (2\lambda_f^2 v)=0,
\end{equation}
where the sum over scalars includes only those from physics beyond the SM and the sum over fermions includes both the SM and new physics contributions\footnote{Here, we have used $m_f=\lambda_fv/\sqrt{2}$; this relation is true for fermions which obtain their masses solely from their Yukawa couplings.  However, the form of the relation in Eq. (\ref{eq:grandrelation}) is also valid in the case of mixing with vectorlike fermions.  We thus obtain a form for the fermionic contributions in Eq. (\ref{eq:grandrelation}) analogous to those for scalars.  We note that the new physics contributions can be written completely in terms of couplings of the new particles to $H$, without reference to the new particle masses; in particular, Eq. (\ref{eq:grandrelation}) does not depend on the scalar masses.}.  We now compare Eq. (\ref{eq:grandrelation}) to the relation for cancellation of quadratic divergences in all Higgs tadpoles (for the analogous relation in the case of only SM fields, see \cite{Veltman:1980mj}),
\begin{equation}
\label{eq:hierarchy}
\left(\frac{6M_W^2}{v}+\frac{m_H^2}{v}\right) + \left(\frac{3M_Z^2}{v}+\frac{m_H^2}{2v}\right)+\left(\frac{3m_H^2}{2v}\right)+\displaystyle\sum_{\rm scalars}  (2\lambda_S v)-\displaystyle\sum_{\rm fermions}(2\lambda_f^2v)=0,
\end{equation} 
where the terms in the second and third sets of parentheses are the contributions from loops of the $Z$ boson (and its associated Goldstone and ghost fields) and the physical Higgs boson, and where we have re-written the terms in the first set of parentheses using $M_W=gv/2$. 

We next note that Eq. (\ref{eq:grandrelation}) can be obtained from Eq. (\ref{eq:hierarchy}) by re-weighting each term in Eq. (\ref{eq:hierarchy}) by the square of the charge of the particle in the loop; hence, terms corresponding to $Z$ and Higgs loops do not appear in Eq. (\ref{eq:grandrelation}).  Next, we note that Eq. (\ref{eq:hierarchy}) is also the relation which must be fulfilled in order for quadratic divergences in the Higgs self-energy to cancel.  This is due to the renormalizeability of the Higgs potential; as both $m_H$ and $v$ are determined by the parameters $\mu^2$ and $\lambda$ in the Higgs potential, a cancellation of quadratic divergences in these parameters simultaneously cancels the quadratic divergences in $m_H$ and $v$.
 
Although Eqs. (\ref{eq:grandrelation}) and (\ref{eq:hierarchy}) are not equivalent, together they have interesting consequences.  Consider the case where we have a model which cancels the quadratic divergences in the Higgs self-energy by the addition of new scalars and/or fermions.  Let us assume that it does this separately for each value of electric charge, i.e., quadratic divergences from neutral particles cancel, quadratic divergences from loops of charge $+1/3$ cancel, etc.  The above results imply that in such a model, the $d=4$ calculation of $H\rightarrow\gamma\gamma$ will be gauge-invariant.  Alternatively, if we (for some reason) assume that the $d=4$ $H\rightarrow\gamma\gamma$ calculation is physically correct, we have a constraint on the particle content of the theory,  Eq. (\ref{eq:grandrelation}).

As the quadratic divergences which contribute to $m_H$ cancel in supersymmetric models, and as supersymmetry accomplishes this by introducing, for each SM particle, a new particle of equal charge, we expect that  Eq. (\ref{eq:grandrelation}) is satisified in the MSSM.  We have checked explicitly and found this to be the case for arbitrary sfermion left-right and flavor mixing and for arbitrary chargino mixing resulting from soft-breaking terms.  Although we do not give all the details here, we will point out one interesting feature of the calculation; here, we will neglect left-right and flavor mixing for simplicity.  The coefficient of the quadratic divergence which would enter Eq. (\ref{eq:grandrelation}) for an $H_0$ tadpole diagram containing an up-type squark $\tilde{u}$ loop is \cite{Gunion:1989we} 
\begin{equation}
\label{eq:susycoup}
\mbox{quad. div}=e_u^2\left[\frac{gM_Z}{\cos\theta_W}(I_u\mp e_u\sin^2\theta_W)\cos(\alpha+\beta)+\frac{gm^2_u}{M_W\sin\beta}\sin\alpha\right],
\end{equation}
where $\theta_W$ is the weak mixing angle, $I_u$ is the isospin ($=1/2$ for $\tilde{u}_L$, $=0$ for $\tilde{u}_R$), $e_u$ is the squark charge,  $m_u$ is the mass of the corresponding fermion, and the $-$ sign holds for $\tilde{u}_L$ while the $+$ holds for $\tilde{u}_R$.  When we add the terms for the $\tilde{u}_L$ and $\tilde{u}_R$ loops, the terms proportional to $e_u\sin^2\theta_W$ cancel.  The terms proportional to $m_u^2$ will cancel with the corresponding fermion tadpoles.  This leaves a leftover term proportional to $e_u^2I_u$.  We note, however, that such terms will happen for each sfermion loop contribution; when contributions from all sfermion loops are summed, $\sum e_f^2 I_f=0$ by the usual triangle anomaly conditions of the SM.   A similar result holds for the $h_0$ tadpoles.  

We will also make a few brief points relevant to models which have been presented as solutions to the hierarchy problem.  First, we point out that not all models which cancel the quadratic divergences in the Higgs self-energy will satisfy Eq. (\ref{eq:grandrelation}); for example, the (now ruled out) scenario where the contributions of SM particles to the quadratic divergences in the Higgs self-energy cancel amongst themselves \cite{Veltman:1980mj} does not satisfy Eq. (\ref{eq:grandrelation}).   Second, although we do not pursue this avenue here, we expect that a relation similar to Eq. (\ref{eq:grandrelation}) would also hold for $H\rightarrow gg$ and thus place a constraint on the colored particle content of the theory; for the case of the MSSM, this cancellation would involve the $SU(3)^2\times U(1)$ and $SU(3)^2\times SU(2)$ anomaly cancellation conditions, similar to those utilized for $H\rightarrow\gamma\gamma$ above.  

\section{Discussion}
\label{sec:dis}
We have argued that assuming the validity of the $d=4$ calculation in $H\rightarrow\gamma\gamma$ has surprisingly interesting consequences.  Not only is it possible to arrange for the gauge-invariance-violating terms to cancel, but also this cancellation is very closely related to the cancellations of quadratic divergences in the Higgs tadpole and self-energy.  In some sense, it could even be loosely argued that the ambiguous integrals in $H\rightarrow\gamma\gamma$ are a side-effect of the hierarchy problem.  As solving the hierarchy problem has been a prime motivation for developing new physics models, it is not surprising that some models already on the market give sensible, gauge-invariant results for $H\rightarrow\gamma\gamma$ calculated in $d=4$.  Additionally, given the close relation between gauge invariance in $H\rightarrow\gamma\gamma$ and the quadratic divergences relevant to the hierarchy problem, our results make it somewhat nonintuitive that one should use dimensional regularization for the calculation of $H\rightarrow\gamma\gamma$ and yet use a regulator that preserves the quadratic divergences in $d=4$ when considering the Higgs tadpole and self-energy; we do not attempt to address this last issue but merely note its nonintuitive nature.

We will briefly mention a few possible extensions of this work.  First, we note that our calculation here was limited to one-loop order; it would be interesting to know if similar behavior holds with higher-order diagrams.  Second, we have not investigated whether or not such relations hold for final states with massive gauge bosons, such as $H\rightarrow ZZ$, $W^+W^-$, $Z\gamma$.  Lastly, we also point out that the regulator dependence of finite integrals is not limited to processes involving scalars; similar behavior also shows up in photon scattering \cite{Khare:1976zx,Liang:2011sj}, a pure QED process.  In the case of photon scattering, however, the application of the Ward identity to two of the external photons yields expressions in terms of the quadratically-divergent photon self-energy; unlike the quadratic divergences in the Higgs case, those in the photon self-energy are customarily treated with dimensional regularization, and considered to not be of physical importance.  Investigating the regulator dependence in photon scattering and possibly seeing if other processes display this behavior are left for future work.

It is interesting to speculate on the possible implications of learning that the relation in Eq. (\ref{eq:grandrelation}) does or does not hold in nature.  Although Eq. (\ref{eq:grandrelation}) is not equivalent to the relation for cancellation of the quadratic divergences in the Higgs self-energy, Eq. (\ref{eq:hierarchy}), their similarity may let one optimistically hope that they may be simultaneously empirically confirmed or refuted.  As other considerations indicate that the hierarchy problem may be solved at the weak scale, and as it is hoped that LHC will thus address this issue, it is possible that we may have a good idea within a few years whether or not Eq. (\ref{eq:grandrelation}) does, in fact, hold.  If it does hold, however, the significance of this result is somewhat unclear.  It may indicate that regulator-dependence of finite calculations is an artifact of not knowing the full theory of nature and should be taken as a clue to unknown physics.  More conservatively, however, it indicates that we should consider the validity of $d=4$ results in other finite but regulator-dependent calculations, such as photon scattering.

\section{Acknowledgements}
The authors would like to thank P. Agrawal, W. Altmannshofer, W. Bardeen, S. Dawson, R. Harnik, J. Lykken, W. Marciano, K. Matchev, T. McElmurry, F. Petriello, P. Ramond, M. Ramsey-Musolf, P. Sikivie, C. Sturm, C. Thorn, J. von Wimmersperg, and F. Yu for their helpful comments and extended discussions.  This work is supported in part under US DOE contract No. DE-FG02-91ER40684 (Northwestern) and US DOE contract No. DE-AC02-07CH11359 (FNAL).  R.V.M. is supported by the Fermilab Graduate Student Fellowship program. 

\appendix
\section{Behavior of tadpole under shifts of loop momentum}
\label{appendix:appa}
Here, we derive Eq. (\ref{eq:quadshift}).  Most of this derivation closely follows a similar derivation in \cite{Elias:1982sq}.

We begin by writing the four quadratically-divergent terms as a difference of linearly-divergent terms,
\begin{align}
\label{eq:appa1}
&\int\frac{d^dp}{(2\pi)^d}\left(\frac{1}{p^2-m^2}-\frac{1}{(p+q_1)^2-m^2}-\frac{1}{(p+q_2)^2-m^2}+\frac{1}{(p+q_1+q_2)^2-m^2}\right)\nonumber\\&=\int\frac{d^dp}{(2\pi)^d}\left[\left(\frac{1}{p^2-m^2}-\frac{1}{(p+q_1)^2-m^2}\right)\right.\nonumber\\&\left.-\left(\frac{1}{(p+q_2)^2-m^2}-\frac{1}{(p+q_1+q_2)^2-m^2}\right)\right]\\&=\int\frac{d^dp}{(2\pi)^d}\left[\left(\frac{2p\cdot q_1+q_1^2}{(p^2-m^2)((p+q_1)^2-m^2)}\right)-\left(\frac{2(p+q_2)\cdot q_1+q_1^2}{((p+q_2)^2-m^2)((p+q_1+q_2)^2-m^2)}\right)\right]\nonumber\\&=\int\frac{d^dp}{(2\pi)^d}\left[\left(\frac{2p\cdot q_1}{(p^2-m^2)((p+q_1)^2-m^2)}\right)-\left(\frac{2(p+q_2)\cdot q_1}{((p+q_2)^2-m^2)((p+q_1+q_2)^2-m^2)}\right)\right],\nonumber
\end{align}
where we have dropped the two terms proportional to $q_1^2$; as these two terms are logarithmically divergent and differ only by a redefintion $p\rightarrow p+q_2$, they cancel.  Next, we rewrite
\begin{align}
\label{eq:appa2}
\frac{p_{\mu}}{(p^2-m^2)((p+q_1)^2-m^2)}=&\frac{p_{\mu}}{(p^2-m^2)}\left[\frac{1}{(p+q_1)^2-m^2}-\frac{1}{p^2-m^2}\right]+\frac{p_{\mu}}{(p^2-m^2)^2}\nonumber\\=&\frac{p_{\mu}(-2p\cdot q_1-q_1^2)}{(p^2-m^2)^2((p+q_1)^2-m^2)}+\frac{p_{\mu}}{(p^2-m^2)^2},
\end{align}
and, similarly,
\begin{align}
\label{eq:appa3}
&\frac{(p+q_2)_{\mu}}{((p+q_2)^2-m^2)((p+q_1+q_2)^2-m^2)}\nonumber\\&=\frac{(p+q_2)_{\mu}}{((p+q_2)^2-m^2)}\left[\frac{1}{(p+q_1+q_2)^2-m^2}-\frac{1}{(p+q_2)^2-m^2}\right]+\frac{(p+q_2)_{\mu}}{((p+q_2)^2-m^2)^2}\\&=\frac{(p+q_2)_{\mu}(-2(p+q_2)\cdot q_1-q_1^2)}{((p+q_2)^2-m^2)^2((p+q_1+q_2)^2-m^2)}+\frac{(p+q_2)_{\mu}}{((p+q_2)^2-m^2)^2}.\nonumber
\end{align}
We thus note that the expressions in Eqs. (\ref{eq:appa2}) and (\ref{eq:appa3}) consist of two pieces, one of which (for $d=4$) is logarithmically divergent and one of which is linearly divergent.  We also note that the expressions in  Eqs. (\ref{eq:appa2}) and (\ref{eq:appa3}) differ by only a momentum redefinition, $p\rightarrow p+q_2$.  We now substitute these last two relations back into Eq. (\ref{eq:appa1}); the logarithmically-divergent terms cancel, and we obtain
\begin{align}
\label{eq:appa4}
&\int\frac{d^dp}{(2\pi)^d}\left(\frac{1}{p^2-m^2}-\frac{1}{(p+q_1)^2-m^2}-\frac{1}{(p+q_2)^2-m^2}+\frac{1}{(p+q_1+q_2)^2-m^2}\right)\nonumber\\&=2q_1^{\mu}\int\frac{d^dp}{(2\pi)^d}\left[\frac{p_{\mu}}{(p^2-m^2)^2}-\frac{(p+q_2)_{\mu}}{((p+q_2)^2-m^2)^2}\right].
\end{align}
We then combine the two terms proportional to $p_{\mu}$ using a Feynman parameter relation, $a^{-2}-b^{-2}=2\int^1_0 dz (b-a)/(az+b(1-z))^3$,
\begin{align}
\label{eq:appa5}
&2q_1^{\mu}\int\frac{d^dp}{(2\pi)^d}\left[\frac{p_{\mu}}{(p^2-m^2)^2}-\frac{(p+q_2)_{\mu}}{((p+q_2)^2-m^2)^2}\right]\nonumber\\&=2q_1^{\mu}\int\frac{d^dp}{(2\pi)^d}\left[\frac{-q_{2\mu}}{((p+q_2)^2-m^2)^2} + \int_0^1 dz \frac{2p_{\mu}(2p\cdot q_2+q_2^2)}{(p^2+(2p\cdot q_2+q_2^2)(1-z)-m^2)^3} \right].
\end{align}
Both integrals in Eq. (\ref{eq:appa5}) are logarithmically divergent, so we can perform a momentum shift.  In the first term, we take $p+q_2\rightarrow p$; in the second, we take $p+(1-z)q_2\rightarrow p$.  These substitutions yield
\begin{align}
\label{eq:appa6}
&2q_1^{\mu}\int\frac{d^dp}{(2\pi)^d}\left[\frac{-q_{2\mu}}{(p^2-m^2)^2} + \int_0^1 dz 2 \frac{2p\cdot q_2 p_{\mu}+(1-z)(1-2z)q_2^2q_{2\mu}}{(p^2+q_2^2z(1-z)-m^2)^3} \right]\nonumber\\=&2q_1^{\mu}\int\frac{d^dp}{(2\pi)^d}\left[\frac{-q_{2\mu}}{(p^2-m^2)^2} + \int_0^1 dz 2 \frac{2p\cdot q_2 p_{\mu}-z(1-2z)q_2^2q_{2\mu}}{(p^2+q_2^2z(1-z)-m^2)^3} \right],
\end{align}
where in the second line we have dropped a term which is odd in $z\leftrightarrow (1-z)$.  Integrating by parts with respect to $z$, this becomes
\begin{align}
\label{eq:appa7}
2q_1^{\mu}\int\frac{d^dp}{(2\pi)^d}&\left[\frac{-q_{2\mu}}{(p^2-m^2)^2} + \int_0^1 dz  \frac{4p\cdot q_2 p_{\mu}}{(p^2+q_2^2z(1-z)-m^2)^3}\right.\nonumber\\ &\left.+ \frac{q_{2\mu}}{(p^2-m^2)^2}+\int^1_0 dz \frac{-q_{2\mu}}{(p^2+q_2^2z(1-z)-m^2)^2}\right]\\=2q_1^{\mu}\int\frac{d^dp}{(2\pi)^d}&\left[\int_0^1 dz  \frac{4p\cdot q_2 p_{\mu}-q_{2\mu}(p^2+q_2^2z(1-z)-m^2)}{(p^2+q_2^2z(1-z)-m^2)^3}\right]\nonumber\\=2q_1^{\mu}q_2^{\nu}\int\frac{d^dp}{(2\pi)^d}&\left[\int_0^1 dz  \frac{4p_{\nu} p_{\mu}-g_{\mu\nu}(p^2+q_2^2z(1-z)-m^2)}{(p^2+q_2^2z(1-z)-m^2)^3}\right]\nonumber.
\end{align}
We now examine Eq. (\ref{eq:appa7}) separately for the cases of $d=4-\epsilon$ and $d=4$.  In the case of $d=4-\epsilon$, the individual terms in Eq. (\ref{eq:appa7}) are finite, and we can reverse the order of integration,
\begin{align}
\label{eq:appa8}
&\int\frac{d^{4-\epsilon}p}{(2\pi)^{4-\epsilon}}\left(\frac{1}{p^2-m^2}-\frac{1}{(p+q_1)^2-m^2}-\frac{1}{(p+q_2)^2-m^2}+\frac{1}{(p+q_1+q_2)^2-m^2}\right)\nonumber\\&=2q_1^{\mu}q_2^{\nu}\int_0^1 dz\int\frac{d^{4-\epsilon}p}{(2\pi)^{4-\epsilon}} \frac{4p_{\nu} p_{\mu}-g_{\mu\nu}(p^2+q_2^2z(1-z)-m^2)}{(p^2+q_2^2z(1-z)-m^2)^3}.
\end{align}
We now see that Eq. (\ref{eq:appa8}) is of the same form as Eq. (\ref{eq:ambdr}), which we see is independent of $m$.  This implies that the integral over $d^{4-\epsilon}p$ is independent of $z$.  Thus, we can replace the momentum integral with its expression for $z=0$; the integral over $z$ is then trivial, and we obtain
\begin{align}
\label{eq:appa9}
&\int\frac{d^{4-\epsilon}p}{(2\pi)^{4-\epsilon}}\left(\frac{1}{p^2-m^2}-\frac{1}{(p+q_1)^2-m^2}-\frac{1}{(p+q_2)^2-m^2}+\frac{1}{(p+q_1+q_2)^2-m^2}\right)\nonumber\\&=2q_1^{\mu}q_2^{\nu}\int\frac{d^{4-\epsilon}p}{(2\pi)^{4-\epsilon}} \frac{4p_{\nu} p_{\mu}-g_{\mu\nu}(p^2-m^2)}{(p^2-m^2)^3}
\end{align}
which is the desired result.

In the case $d=4$, we perform the substitution $4p_{\mu}p_{\nu}\rightarrow g_{\mu\nu}p^2$ on Eq. (\ref{eq:appa7}).  This yields a finite integral; again, we reverse the order of integration and obtain
\begin{align}
\label{eq:appa10}
&\int\frac{d^{4}p}{(2\pi)^{4}}\left(\frac{1}{p^2-m^2}-\frac{1}{(p+q_1)^2-m^2}-\frac{1}{(p+q_2)^2-m^2}+\frac{1}{(p+q_1+q_2)^2-m^2}\right)\nonumber\\&=2q_1^{\mu}q_2^{\nu}\int_0^1 dz\int\frac{d^{4}p}{(2\pi)^{4}} \frac{-g_{\mu\nu}(+q_2^2z(1-z)-m^2)}{(p^2+q_2^2z(1-z)-m^2)^3}.
\end{align}
Similar to the $d=4-\epsilon$ case, we compare this integral to that in Eq. (\ref{eq:amb4d}) and see that the momentum integral is independent of $z$.  Thus, we replace the momentum integral in Eq. (\ref{eq:appa10}) with its value for $z=0$, which also renders the $z$ integral trivial.  Thus,
\begin{align}
\label{eq:appa11}
&\int\frac{d^{4}p}{(2\pi)^{4}}\left(\frac{1}{p^2-m^2}-\frac{1}{(p+q_1)^2-m^2}-\frac{1}{(p+q_2)^2-m^2}+\frac{1}{(p+q_1+q_2)^2-m^2}\right)\nonumber\\&=2q_1^{\mu}q_2^{\nu}\int\frac{d^{4}p}{(2\pi)^{4}} \frac{-g_{\mu\nu}(-m^2)}{(p^2-m^2)^3}.  
\end{align}
Then, as $\int\frac{d^{4}p}{(2\pi)^{4}} (4p_{\mu}p_{\nu}-p^2g_{\mu\nu})/(p^2-m^2)^3=0$ for $d=4$, we can write this as
\begin{align}
\label{eq:appa12}
&\int\frac{d^{4}p}{(2\pi)^{4}}\left(\frac{1}{p^2-m^2}-\frac{1}{(p+q_1)^2-m^2}-\frac{1}{(p+q_2)^2-m^2}+\frac{1}{(p+q_1+q_2)^2-m^2}\right)\nonumber\\&=2q_1^{\mu}q_2^{\nu}\int\frac{d^{4}p}{(2\pi)^{4}} \frac{4p_{\mu}p_{\nu}-g_{\mu\nu}(p^2-m^2)}{(p^2-m^2)^3},  
\end{align}
which is the desired result.

\section{Details of $W$ loop calculation}
\label{appendix:appb}
Here, we sum the contributions from loops involving $W$ bosons, Goldstone bosons, and ghosts shown in Fig. \ref{fig:wdiag} for $q_1=q_2=p_H=0$.  We will not consider those terms from diagrams containing only scalars (diagrams c) and d)), as their treatment is identical to that covered in Sec. \ref{sec:fands}.  We will also neglect the contribution of diagram f), which, although nonzero, is finite and, therefore, regulator-independent.  First, we list the results for each of these diagrams, removing, for simplicity, a common factor of $\varepsilon^{*\mu}_1\varepsilon^{*\nu}_2(e^2 g M_W)$; we will label each of these terms according to their designation (a-j) in Fig. \ref{fig:wdiag}.  These terms are:  
\begin{align}
\label{allwterms}
&a= \int \frac{d^dp}{(2\pi)^d}  \frac{2}{(p^2-M_W^2)^3}\times\nonumber\\&\left[\vphantom{\frac{(1-\xi)}{(p^2-\xi M_W^2)}} \left(2p^2 g_{\mu\nu} + (10-4\epsilon) p_{\mu}p_{\nu}\right)- \frac{(1-\xi)}{(p^2-\xi M_W^2)}(3p^4 g_{\mu\nu} -3 p^2 p_{\mu}p_{\nu})  + \frac{(1-\xi)^2}{(p^2-\xi M_W^2)^2}(p^6 g_{\mu\nu} -p^4 p_{\mu}p_{\nu})\right]\nonumber,\\
&b=\int \frac{d^dp}{(2\pi)^d}  \frac{-1}{(p^2-M_W^2)^2}\times\nonumber\\&\left[(6-2\epsilon)g_{\mu\nu}-2\frac{(1-\xi)}{(p^2-\xi M_W^2)}(2 p^2 g_{\mu\nu} - 2p_{\mu}p_{\nu}) +\frac{(1-\xi)^2}{(p^2-\xi M_W^2)^2}(2p^4g_{\mu\nu}-2p^2p_{\mu}p_{\nu})\right],\nonumber\\
&e=\int \frac{d^dp}{(2\pi)^d}  (-2\xi)\frac{p_{\mu}p_{\nu}}{(p^2-\xi M_W^2)^3},\nonumber\\
&g=\int \frac{d^dp}{(2\pi)^d}  \frac{4}{(p^2-\xi M_W^2)^2} \frac{1}{(p^2- M_W^2)}  \left[ p_{\mu}p_{\nu} -(1-\xi)\frac{p^2 p_{\mu}p_{\nu}}{(p^2-\xi M_W^2)}\right],\\
&h=\int \frac{d^dp}{(2\pi)^d} (-2\xi) \frac{1}{(p^2-\xi M_W^2)^2} \frac{1}{(p^2- M_W^2)} \left[p^2 g_{\mu\nu} -p_{\mu}p_{\nu} \right],\nonumber\\
&i=\int \frac{d^dp}{(2\pi)^d}  \frac{-2M_W^2}{(p^2-\xi M_W^2)} \frac{1}{(p^2- M_W^2)^2}\left[g_{\mu\nu} -(1-\xi)\frac{2p_{\mu}p_{\nu}}{(p^2-\xi M_W^2)} + (1-\xi)^2 \frac{p^2p_{\mu}p_{\nu}}{(p^2-\xi M_W^2)^2}\right],\nonumber\\
&j=\int \frac{d^dp}{(2\pi)^d}  \frac{-2}{(p^2-\xi M_W^2)} \frac{1}{(p^2- M_W^2)} \left[g_{\mu\nu} -(1-\xi) \frac{p_{\mu}p_{\nu}}{(p^2-\xi M_W^2)} \right].\nonumber
\end{align}
Next, we will combine two sets of these terms,
\begin{align}
a+b=\int \frac{d^dp}{(2\pi)^d} \frac{1}{(p^2-M_W^2)^3}\times&\left[\vphantom{\frac{(1-\xi)}{(p^2-\xi M_W^2)}}((2\epsilon-2)p^2+(6-2\epsilon)M_W^2)g_{\mu\nu}+(20-8\epsilon)p_{\mu}p_{\nu}\right.\nonumber\\& +\frac{(1-\xi)}{(p^2-\xi M_W^2)} (-2p^2-4M_W^2)(p^2g_{\mu\nu}-p_{\mu}p_{\nu})\\&\left.+ \frac{(1-\xi)^2}{(p^2-\xi M_W^2)^2}2M_W^2(p^4 g_{\mu\nu} -p^2 p_{\mu}p_{\nu})\right],\nonumber
\end{align}
and
\begin{align}
e+g=\int \frac{d^dp}{(2\pi)^d} \frac{(2\xi) p_{\mu}p_{\nu}}{(p^2-\xi M_W^2)^3}.
\end{align}
Next, we note that $i$ is entirely finite, and, thus, we can make the substitution $p_{\mu}p_{\nu}\rightarrow g_{\mu\nu}p^2/4$:
\begin{equation}
i=\int \frac{d^dp}{(2\pi)^d}  \frac{-2M_W^2}{(p^2-\xi M_W^2)} \frac{1}{(p^2- M_W^2)^2}\left[1 -\frac{(1-\xi)p^2/2}{(p^2-\xi M_W^2)} +  \frac{(1-\xi)^2p^4/4}{(p^2-\xi M_W^2)^2}\right]g_{\mu\nu}.
\end{equation}
We rewrite each of the above integrands such that they all contain a common denominator and perform the substitution $p_{\mu}p_{\nu}\rightarrow g_{\mu\nu}p^2/4$ on finite terms.  For the sums $a+b$ and $e+g$ above, we obtain
\begin{align}
a+b=&\int \frac{d^dp}{(2\pi)^d} \frac{1}{(p^2-M_W^2)^3(p^2-\xi M_W^2)^3}\nonumber\times\\&\left[\vphantom{\frac{(1-\xi)}{(p^2-\xi M_W^2)}}\left(g_{\mu\nu}\left((2\epsilon-2)p^2+(6-2\epsilon)M_W^2\right)+p_{\mu}p_{\nu}(20-8\epsilon)\right)\right.\nonumber\\&\times\left(p^6-3p^4\xi M_W^2+3p^2\xi^2M_W^4-\xi^3M_W^6\right)\nonumber\\& +(1-\xi) (-2p^2-4M_W^2)(p^2g_{\mu\nu}-p_{\mu}p_{\nu})(p^4-2p^2\xi M_W^2+\xi^2M_W^4)\nonumber\\&\left.+ (1-\xi)^22M_W^2(p^4 g_{\mu\nu} -p^2 p_{\mu}p_{\nu})(p^2-\xi M_W^2)\vphantom{\frac{(1-\xi)}{(p^2-\xi M_W^2)}}\right],\nonumber\\=&\int \frac{d^dp}{(2\pi)^d} \frac{1}{(p^2-M_W^2)^3(p^2-\xi M_W^2)^3}\times\\&\left[\vphantom{\frac{(1-\xi)}{(p^2-\xi M_W^2)}}g_{\mu\nu}\left(p^8(2\epsilon-2)+p^6M_W^2(6-9\xi)+p^4M_W^49\xi(-2+\xi)+p^2M_W^63\xi^2(6-\xi)-6\xi^3M_W^8\right)\right.\nonumber\\&+g_{\mu\nu}(1-\xi)\frac{3}{4}\left(-\frac{8}{3}p^8+p^6M_W^2(4\xi-4)+p^4M_W^4(-2\xi^2+8\xi)-p^2M_W^64\xi^2\right)\nonumber\\&+g_{\mu\nu}(1-\xi)^2\left(\frac{3M_W^2}{2}\left(p^6-p^4\xi M_W^2\right)\right)\nonumber\\&\left.+p_{\mu}p_{\nu}p^6(22-8\epsilon-2\xi)\vphantom{\frac{(1-\xi)}{(p^2-\xi M_W^2)}}\right]\nonumber\\=&\int \frac{d^dp}{(2\pi)^d} \frac{1}{(p^2-M_W^2)^3(p^2-\xi M_W^2)^3}\nonumber\times\\&\left[g_{\mu\nu}\left(p^8(2\epsilon-4+2\xi)+p^6M_W^2\left(\frac{9}{2}-6\xi-\frac{3}{2}\xi^2\right)+p^4M_W^4\left(-\frac{27}{2}\xi+\frac{9}{2}\xi^2\right)\right.\right.\nonumber\\&\left.\left.+p^2M_W^6(15\xi^2)+M_W^8(-6\xi^3)\vphantom{\frac{1}{1}}\right)+p_{\mu}p_{\nu}p^6(22-8\epsilon-2\xi)\right]\nonumber,
\end{align}
and
\begin{align}
e+g=&\int \frac{d^dp}{(2\pi)^d} (2\xi)\frac{ p_{\mu}p_{\nu}(p^6-3p^4 M_W^2+3p^2M_W^4-M_W^6)}{(p^2-M_W^2)^3(p^2-\xi M_W^2)^3}\nonumber\\=&\int \frac{d^dp}{(2\pi)^d} (2\xi)\frac{ (p_{\mu}p_{\nu}p^6)+g_{\mu\nu}\frac{1}{4}(-3p^6 M_W^2+3p^4M_W^4-p^2M_W^6)}{(p^2-M_W^2)^3(p^2-\xi M_W^2)^3}.
\end{align}
For the remaining terms, we get
\begin{align}
h=&\int \frac{d^dp}{(2\pi)^d} (-2\xi) \frac{(p^2 g_{\mu\nu} -p_{\mu}p_{\nu})}{(p^2- M_W^2)^3(p^2-\xi M_W^2)^3 } \left(p^6-(2+\xi)p^4M_W^2+(2\xi+1)p^2M_W^4-\xi M_W^6\right)\nonumber\\=&\int \frac{d^dp}{(2\pi)^d} (-2\xi) \times\\&\frac{(p^2 g_{\mu\nu} -p_{\mu}p_{\nu})p^6+g_{\mu\nu}\frac{3}{4}(-(2+\xi)p^6M_W^2+(2\xi+1)p^4M_W^4-\xi p^2M_W^6)}{(p^2- M_W^2)^3(p^2-\xi M_W^2)^3 }\nonumber,
\end{align}
and
\begin{align}
i=&\int \frac{d^dp}{(2\pi)^d}  \frac{-2M_W^2}{(p^2- M_W^2)^3(p^2-\xi M_W^2)^3}g_{\mu\nu}\times\nonumber\\&\left[\vphantom{\frac{(1-\xi)}{2}}(p^6-p^4M_W^2(1+2\xi)+p^2M_W^4(\xi^2+2\xi)-\xi^2M_W^6) \right.\nonumber\\&\left.-\frac{(1-\xi)}{2}(p^6-(1+\xi) p^4M_W^2+\xi p^2M_W^4) + \frac{(1-\xi)^2}{4}(p^6-p^4M_W^2)\right]\\=&\int \frac{d^dp}{(2\pi)^d}  \frac{-2M_W^2}{(p^2- M_W^2)^3(p^2-\xi M_W^2)^3}g_{\mu\nu}\times\nonumber\\&\left[p^6\left(\frac{3+\xi^2}{4}\right)+p^4M_W^2\left(\frac{-3-6\xi-3\xi^2}{4}\right)+p^2M_W^4\left(\frac{3(\xi^2+\xi)}{2}\right)-\xi^2M_W^6\right],\nonumber
\end{align}
and
\begin{align}
j=&\int \frac{d^dp}{(2\pi)^d}  \frac{-2}{(p^2- M_W^2)^3(p^2-\xi M_W^2)^3} \times\nonumber\\&\left[g_{\mu\nu}\left(p^8+p^6M_W^2(-2-2\xi)+p^4M_W^4(1+4\xi+\xi^2)+p^2M_W^6(-2\xi-2\xi^2)+\xi^2M_W^8\right)\right.\nonumber\\& \left.-(1-\xi) p_{\mu}p_{\nu}\left(p^6-p^4M_W^2(2+\xi)+p^2M_W^4(1+2\xi)-\xi M_W^6\right) \right]\nonumber\\=&\int \frac{d^dp}{(2\pi)^d}  \frac{-2}{(p^2- M_W^2)^3(p^2-\xi M_W^2)^3} \times\\&\left[g_{\mu\nu}\left(p^8+p^6M_W^2\left(\frac{-6-9\xi-\xi^2}{4}\right)+p^4M_W^4\left(\frac{3+15\xi+6\xi^2}{4}\right)\right.\right.\nonumber\\ &\quad\quad\left.\left.+p^2M_W^6\left(\frac{-7\xi-9\xi^2}{4}\right) +\xi^2M_W^8\right)-p_{\mu}p_{\nu}p^6(1-\xi)\right].\nonumber
\end{align}
Summing all of these terms,
\begin{align}
\label{eq:atoj}
&a+b+e+g+h+i+j=\int \frac{d^dp}{(2\pi)^d}  \frac{1}{(p^2- M_W^2)^3(p^2-\xi M_W^2)^3} \times\nonumber\\&\left[g_{\mu\nu}\left(p^8\left(2\epsilon-6\right)+p^6M_W^2\left(6\right)+p^4M_W^4\left(-18\xi\right)+p^2M_W^6\left(18\xi^2\right)+M_W^8\left(-6\xi^3\right)\right)\right.\\&\left.+p_{\mu}p_{\nu}p^6\left(24-8\epsilon\right)\right].\nonumber
\end{align}
We now note that 
\begin{align}
&\int \frac{d^dp}{(2\pi)^d}  \frac{4p_{\mu}p_{\nu}-g_{\mu\nu}(p^2-M_W^2)}{(p^2- M_W^2)^3}\nonumber\\=&\int \frac{d^dp}{(2\pi)^d}  \frac{4p_{\mu}p_{\nu}-g_{\mu\nu}(p^2-M_W^2)}{(p^2- M_W^2)^3(p^2-\xi M_W^2)^3}(p^2-\xi M_W^2)^3\\=&\int \frac{d^dp}{(2\pi)^d}  \frac{4p_{\mu}p_{\nu}-g_{\mu\nu}(p^2-M_W^2)}{(p^2- M_W^2)^3(p^2-\xi M_W^2)^3}(p^6-3p^4\xi M_W^2+3p^2\xi^2M_W^4-\xi^3M_W^6)\nonumber\\=&\int \frac{d^dp}{(2\pi)^d}  \frac{4p^6p_{\mu}p_{\nu}-g_{\mu\nu}p^8+g_{\mu\nu}M_W^2(p^6-3p^4\xi M_W^2+3p^2\xi^2M_W^4-\xi^3M_W^6)}{(p^2- M_W^2)^3(p^2-\xi M_W^2)^3},\nonumber
\end{align}
where in the last line we have substituted $4p_{\mu}p_{\nu}\rightarrow p^2g_{\mu\nu}$.  Comparing with (\ref{eq:atoj}) and dropping terms where $\epsilon$ multiplies something purely finite, we obtain
\begin{align}
a+b+e+g+h+i+j=(6-2\epsilon)\int \frac{d^dp}{(2\pi)^d}  \frac{4p_{\mu}p_{\nu}-g_{\mu\nu}(p^2-M_W^2)}{(p^2- M_W^2)^3},
\end{align}
and, thus,
\begin{align}
i{\cal M}_{\mu\nu}^{a,b,e,g-j}\varepsilon^{*\mu}_1\varepsilon^{*\nu}_2=\varepsilon^{*\mu}_1\varepsilon^{*\nu}_2 e^2 g M_W (6-2\epsilon)\int \frac{d^dp}{(2\pi)^d}  \frac{4p_{\mu}p_{\nu}-g_{\mu\nu}(p^2-M_W^2)}{(p^2- M_W^2)^3},
\end{align}
which yields the relevant terms in Eq. (\ref{eq:appbsum}).

\bibliographystyle{h-physrev}

\end{document}